\documentclass[a4paper]{jpconf}
\usepackage{graphicx}
\begin{document}
\begin{flushleft}
INR-TH-2016-010
\end{flushleft}

\title{From perturbative calculations of the QCD static potential
towards four-loop pole-running heavy quarks  masses relation}

\author{A. L. Kataev$^{1,2}$, V. S. Molokoedov$^{1,2}$ }

\address{$^1$ Institute for Nuclear Research of the Russian Academy of Sciences (INR), 60th October Anniversary Prospect, 7a, 117312 Moscow, Russia }
\address{$^2$ Moscow Institute of Physics and Technology (MIPT), 141700, Dolgoprudny, Russia}

\ead{kataev@ms2.inr.ac.ru$^1$, viktor\_molokoedov@mail.ru$^2$ }

\begin{abstract}
The summary  of the available semi-analytical  
results for the  three-loop   corrections to the QCD static 
potential and for the   $\mathcal{O}(\alpha_s^4)$ contributions to the ratio 
of the running and pole heavy quark masses   
are  presented.  
The procedure of the determination of 
the dependence of the four-loop contribution to the 
pole-running heavy quarks mass ratio 
on  the number of 
quarks flavours, based on application of the  least squares 
method is described. The necessity of clarifying the reason of discrepancy
between the numerical uncertainties of the $\alpha_s^4$ coefficients
in the mass ratio, obtained by  this mathematical method by the direct 
numerical calculations is emphasised. 
\end{abstract}

\section{Introduction.}

It is known that in the  Standard Model it is possible to introduce 
several definitions of  heavy quarks 
masses. Rather  applicable at present are the 
potential subtracted masses $m_{PS,q}(\mu^2)$  \cite{Beneke:1998rk}, 
the $\rm\overline{MS}$-scheme running  
masses $\overline{m}_q(\mu^2)$ and the scale-independent pole masses 
$M_{q}$. The relations between these  definitions of heavy quarks 
masses were  studied  in the number 
of theoretical works on the subject (see e.g. \cite{Beneke:1998rk},
\cite{Ayala:2014yxa}, \cite{Marquard:2015qpa}).    
Since  these  masses are extracted at present  
from different  rather precise 
experimental data,  e.g. from  the spectroscopy of heavy hadrons  
and from the measurable  cross-sections of the 
productions of heavy quarks in different observable processes, 
it is important to know the relations between $M_{q}$,  $\overline{m}_q(\mu^2)$
and  $m_{PS,q}(\mu^2)$ with high enough precision, which is 
gained  by evaluation   of   high order  perturbative 
corrections to the relations between pole and $\rm\overline{MS}$-scheme 
running heavy quarks masses and to the QCD static potential as well. 
This talk is devoted to 
the summary  of the available 
results of the  determination of three-loop corrections to the QCD static 
potential and to  the $\mathcal{O}(\alpha_s^4)$ relations between different 
definitions of heavy quarks masses. Special attention 
is paid to the discussions  of theoretical 
errors in  the existing  numerical calculations of the  $\mathcal{O}(\alpha_s^4)$ 
contributions to the  static potential \cite{Smirnov:2008pn},\cite{Smirnov:2009fh},\cite{Anzai:2009tm} and to 
the ratio of the running and pole heavy quark masses
\cite{Marquard:2015qpa}. The  determination of the dependence of the 
this QCD correction on  the number of lighter quarks flavours $n_l$ by means 
of the ordinary  least squares (OLS)  mathematical method \cite{Kataev:2015gvt} 
is described.  

\section{The  static potential in perturbative QCD.}

The QCD static  potential is strictly defined through   
the vacuum expectation value of the Wilson loop \cite{Brambilla:2004jw}:
\begin{eqnarray}
\label{V}
{\rm{V}}(r)=-\lim_{T \to \infty}\frac{1}{iT} \ln \frac{\langle 0\mid {\rm{Tr}~ P} e^{ig\oint\limits_C dx^\mu A^a_\mu t^a}\mid 0\rangle}{\langle 0\mid {\rm{Tr}} 1\mid 0\rangle}
 = \displaystyle\int\frac{d^3\vec{q}}{(2\pi)^3}e^{i\vec{q}\vec{r}}\;\tilde{\rm{V}}(\vec{q}^{\; 2})
\end{eqnarray}
where $C$ is a closed rectangular contour, T and $r$ are the time and three-dimensional 
space variables, $P$ is the path-ordering operator, 
$A^a_{\mu}$ is a gluon field and $t^a$ are the   generators of Lie algebra of 
the $\rm{SU}(N_c)$ group.

Perturbation theory  expression for the Fourier transform 
$\tilde{\rm{V}}(\vec{q}^{\; 2})$ of the 
static potential is known up to $\mathcal{O}(\alpha^4_s)$-corrections:  
\begin{equation}
\label{tildeV}
\tilde{\rm{V}}(q)=-\frac{4\pi C_F\alpha_s(\vec{q}^{\; 2})}{\vec{q}^{\; 2}}
\left(1+a_1 a_s(\vec{q}^{\;2}) +a_2 a^2_s(\vec{q}^{\;2})
+\left(a_3+ \frac{\pi^2 C^3_A}{8}\ln\frac{\mu^2}{\vec{q}^{\; 2}}\right) 
a^3_s(\vec{q}^{\;2}) \right)
\end{equation}
where $C_F$ and $C_A$ are the   Casimir operators
and the QCD coupling constant $a_s(\vec{q}^{\;2})= \alpha_s(\vec{q}^{\;2})/\pi$
is defined in the $\overline{\rm{MS}}$-scheme. The  coefficients 
$a_1$ and $a_2$  in Eq.~(\ref{tildeV})
were obtained by evaluating in the analytical form   
the corresponding one and two-loops Feynman diagrams  (see \cite{Fischler:1977yf},\cite{Billoire:1979ih} and \cite{Peter:1996ig}, 
\cite{Schroder:1998vy} respectively).  
The additional term $\pi^2C^3_A/8$, correctly evaluated in  
\cite{Kniehl:2002br}, arises due to the infrared (IR)
divergences, which begin to manifest themselves in the
static potential at the three-loop level. However, in the concrete  
applications of the effective non-relativistic QCD these  
IR-divergent terms are  cancelling out. 

The three-loop  contribution     $a_3$ to    Eq.~(\ref{tildeV}) can be presented as
\begin{equation}
\label{a3}
a_3=a^{(3)}_3n^3_l+a^{(2)}_3n^2_l+a^{(1)}_3n_l+a^{(0)}_3
\end{equation}
where $n_l$ are the number of quarks, which  
contribute to the QCD corrections  
to the static potential, responsible for the 
strong  interactions of the heavy  
quark-antiquark pair with the flavour number $n_f=n_l+1$.     
Note, that the masses of all lighter  
quarks with the flavour number 
$n_l$ are usually  neglected.  

The $n^3_l$ and $n^2_l$-dependent terms were  evaluated analytically 
in \cite{Smirnov:2008pn} and have the following form
\begin{eqnarray}
\label{a33}
a^{(3)}_3=-\frac{125}{729}T^3_F~, ~~~~~~
a^{(2)}_3=\left(\frac{12541}{15552}+\frac{23}{12}\zeta(3)+\frac{\pi^4}{135}\right)C_AT^2_F
+\left(\frac{7001}{2592}-\frac{13}{6}\zeta(3)\right)C_FT^2_F
\end{eqnarray}
Definite parts of  the  
$n_l$-dependent coefficient $a^{(1)}_3$ and the overall 
expression for the   constant term $a^{(0)}_3$   
have not yet been computed analytically. 
The semi-analytical expression for $a_3^{(1)}$ was 
obtained  in \cite{Smirnov:2008pn} and reads: 
\begin{eqnarray}
\label{a13}
a^{(1)}_3&=&-\frac{709.717}{64} C^2_AT_F
+\left(-\frac{71281}{10368}+\frac{33}{8}\zeta(3)+\frac{5}{4}\zeta(5)\right)C_AC_FT_F  \\ \nonumber
&+&\left(\frac{143}{288}+\frac{37}{24}\zeta(3)-\frac{5}{2}\zeta(5)\right)C^2_FT_F-
\frac{56.83(1)}{64}\frac{d^{abcd}_Fd^{abcd}_F}{N_A}~~. 
\end{eqnarray}
where $d^{abcd}_F= {\rm{Tr}}(t^at^{(b}t^ct^{d)})/6 \;$ and
$d^{abcd}_A={\rm{Tr}}(C^aC^{(b}C^cC^{d)})/6 \;$ is the   total 
symmetric tensors,  $(C^a)_{bc}=-if^{abc}$ are the generators of the adjoint 
representation of the  Lie algebra of the  $\rm{SU(N_c)}$ group. 
Note that in the QED limit with
$C_A$=0 the $n_l$-dependent
terms, which are proportional to the powers of $T_F$ in (\ref{a33}) and (\ref{a13}), are in agreement with 
the $\overline{\rm{MS}}$-scheme three-loop corrections to 
the photon vacuum polarisation function in QED,  
presented previously  in \cite{Gorishnii:1991hw}.  
The inaccuracy of the  numerical evaluation of the $C^2_AT_F$
coefficient was  not indicated in  \cite{Smirnov:2008pn},  
since  it is much smaller than   the uncertainty  of the 
$d^{abcd}_Fd^{abcd}_F$ coefficient in the presented in (\ref{a13})  
expression for the   $a_3^{(1)}$-term.  
The most precise calculation  for $a_3^{(0)}$  was made  in  
\cite{Smirnov:2009fh}. Its result 
 \begin{equation}
\label{a03}
a^{(0)}_3=\frac{502.24(1)}{64}C^3_A-\frac{136.39(12)}{64}\frac{d^{abcd}_Fd^{abcd}_A}{N_A}
\end{equation}
was published almost simultaneously with the following result of 
the independent calculation, performed in \cite{Anzai:2009tm}:  
\begin{equation}
\label{anzai}
a^{(0)}_3=\frac{502.22(12)}{64}C^3_A-\frac{136.8(14)}{64}\frac{d^{abcd}_Fd^{abcd}_A}{N_A}
\end{equation}
Using the same computer code  one of the authors 
of  \cite{Anzai:2009tm} improved recently the numerical 
precision of the second  term in (\ref{anzai}), which now reads\footnote
{We are grateful to Y. Sumino for informing us about this new unpublished result of his personal calculations.}: 
\begin{equation}
\label{Sumino}
a^{(0)}_3=\frac{502.22(12)}{64}C^3_A-\frac{136.6(2)}{64}\frac{d^{abcd}_Fd^{abcd}_A}{N_A}
\end{equation}
Within the error bars the  results  (\ref{a03}), (\ref{anzai}) and (\ref{Sumino})
have a common intersection region. The   obtained in \cite{Smirnov:2009fh} 
contributions  to  
$a^{(0)}_3$ have  the   smallest  uncertainties. However, it  is highly 
desirable to understand what is the reason of the differences in  
the numerical results of   (\ref{a03}), (\ref{anzai}) and (\ref{Sumino}). 

Indeed, in the process of calculations the authors of  \cite{Smirnov:2009fh} and  \cite{Anzai:2009tm} used the same  theoretical methods. 
Let us  
mention the basic steps of these calculations.  
 First, the 
diagrams, contributing to the  3-loop correction to static potential  
are generated.  Then they are reduced to about   2000 3-loop  integrals.
In  \cite{Smirnov:2009fh} these integrals   
were reduced 
to   a relatively small number  of the concrete  
master integrals using  the integration-by-parts (IBP)  method  
\cite{Chetyrkin:1981qh} and the FIRE algorithm  \cite{Smirnov:2008py}, 
which is based in part on  the    
Laporta algorithm \cite{Laporta:2001dd}.  
Almost all master
integrals were evaluated 
analytically apart of {\it  three} of them, which were 
evaluated numerically. 
To check the results of these  analytical and numerical calculations  
the program FIESTA \cite{Smirnov:2008py},
\cite{Smirnov:2013eza} was used. The obtained in  
\cite{Smirnov:2009fh} and  \cite{Smirnov:2008pn} 
 numerical results in the expressions for $a_3^{(0)}$ and $a_3^{(1)}$ terms 
are related to the results of the numerical evaluation 
of {\it  three} mentioned above master  integrals.

In the independently made work   \cite{Anzai:2009tm} the  
IBP method \cite{Chetyrkin:1981qh}  and the Laporta algorithm 
were applied in the  another  way.  
To evaluate the master integrals the  different from FIESTA computer program, 
written by the authors of  \cite{Anzai:2009tm} 
was used. Therefore, it is difficult to relate the numerical uncertainties of 
the results, given in (\ref{anzai}) and (\ref{Sumino}) with the ones, 
presented in (\ref{a03}), which are determined by the numerical evaluation 
of the known {\it three} master integrals. In view of this it may be 
of real interest to consider  the possibility of the detailed comparison of 
the  FIESTA computer program with the one, created by the authors of 
\cite{Anzai:2009tm}. 

Note, that in \cite{Kataev:2015yha} using the maximal transcendentality 
hypothesis the guess on the analytical structure of    
{\it four} numerically known coefficients in the the expression 
for the $\mathcal{O}(a_s^3)$ correction to the static potential was made.
More definitely, in  \cite{Kataev:2015yha} we 
proposed to consider the reliability   
of the   existence of the  following decomposition 
\begin{eqnarray}
\label{1}
&&709.717=R_{1}+R_2\pi^2+R_3\pi^4+R_4\zeta(3)+R_5\pi^2\zeta(3)+R_6\zeta(5) \\
\label{2}
&&502.24(1)=R_7+R_8\pi^2+R_9\pi^4+R_{10}\zeta(3)+R_{11}\pi^2\zeta(3)+R_{12}\zeta(5) \\
\label{3}
&&56.83(1)=R_{13}+R_{14}\pi^2+R_{15}\pi^4+ R_{16}\zeta(3) \\ \label{4}
&&136.39(12)= R_{17}+R_{18}\pi^2+R_{19}\pi^4+ R_{20}\zeta(3) 
\end{eqnarray}
where $R_i$ are still  unknown rational numbers. 
One should note that the concrete coefficients  
 $R_i$  in (\ref{1})-(\ref{4}) 
may be zero. Considering carefully the analytical results 
of  \cite{Smirnov:2008pn} in   \cite{Kataev:2015yha} we expressed 
the hope that $R_5$, $R_{11}$,  $R_{12}$ and $R_{18}$  may be really zero.
The status of the  discussed  above hypothesis of  \cite{Kataev:2015yha} will 
be clarified when  the remaining numerically calculated  in 
\cite{Smirnov:2008pn} {\it three}   master integrals will be evaluated analytically. The work on 
the solution of this calculation problem is now in progress \footnote{We are grateful to V.A.Smirnov for this information.}.   

\section{The relation between static potential and potential subtracted 
heavy quark masses.} 

The information on the   QCD static potential and on the uncertainties 
of its perturbative calculations  are important in the 
definitions of the  heavy quarks masses and their values. This is possible 
to understand after consideration of  introduced in \cite{Beneke:1998rk}
notion of heavy quark masses  in the potential subtraction scheme. 
This   definition  is  motivated by 
the studies of the behaviour of the cross-sections of the productions of  
heavy quarks  near  thresholds. It  is  based on the  
long-distance modifications of the perturbative static heavy quark 
potential $\tilde{\rm{V}}(\vec{q}^{\;2})$, oriented on
the suppression of the IR renormalon contributions (for the 
consideration  of the possible other power-suppressed non-perturbative 
contributions to the Coulomb-like potential for heavy quarks, which are 
 not related to the  IR renormalon effects, see e.g. \cite{Akhoury:1997by}).    
In the static potential the IR  effects can be eliminated by 
introducing  boundary condition $\vert \vec{q}\vert>\mu_f$ to its   
Fourier transform, where the factorisation scale $\mu_f$  
varies  in the  region   $\Lambda_{QCD}<\mu_f<M_q{\it{v}}$, and ${\it v}$
is the relative velocity of two  heavy quarks, 
which is rather small near thresholds of their production. 
The equivalent way of the fixation of  the factorisation scale  
is motivated by the uncertainty principle, which 
gives  $\mu_f\sim 1/r\sim M_q{\it v}\sim M_q\alpha_s(M_q)$. 
The factorisation scale $\mu_f$ enter  the   
definition 
of the   subtracted potential ${\rm{V}}(r, \mu_f)$  as:
\begin{eqnarray}
\label{Vrm}
{\rm{V}}(r,\mu_f)={\rm{V}}(r)+2\delta m_q(\mu_f)~~~.
\end{eqnarray} 
The  residual heavy quark mass $\delta m_q(\mu_f)$ is determined 
in the following way 
\begin{eqnarray}
\delta m_q(\mu_f)=-\frac{1}{2}\int\limits_{\mid \vec{q} \mid <\mu_f} \frac{d^3\vec{q}}{(2\pi)^3}\tilde{{\rm{V}}}(\vec{q}^{\; 2})
\end{eqnarray}
where the perturbative expression for $\tilde{{\rm{V}}}(\vec{q}^{\; 2})$ is 
presented above  in (\ref{tildeV}). 
The  defined by this way subtracted potential ${\rm{V}}(r, \mu_f)$ 
does not contain long-distance contributions. 
As a consequence,  in the  
calculations of the cross sections of heavy quark production 
in the near threshold region 
the 
pole mass $M_q$ (which  contain long-distance IR renormalon contributions in 
its relation 
to the $\rm{\overline{MS}}$-scheme running mass \cite{Beneke:1994sw}, \cite{Bigi:1994em})  
should be changed to  the potential subtracted quark mass 
$m_{{\rm{PS}},q}(\mu_f)$, defined  in  \cite{Beneke:1998rk}
as 
\begin{eqnarray}
\label{mps}
m_{{\rm{PS}},q}(\mu_f)=M_q-\delta m_q(\mu_f)~~.
\end{eqnarray}
In the r.h.s. of (\ref{mps})  the high loop   perturbative corrections to  
$\delta m_q(\mu_f)$, which generate 
IR renormalon contributions, are 
cancelled exactly with long-distance contributions to the pole mass. 
Using now the expansion of the  pole mass through the   
$\rm{\overline{MS}}$-scheme running   mass
\begin{eqnarray}
M_q=\overline{m}_q(\overline{m}^2_q)(1+ \sum\limits_{i\geq 1} l_ia^i_s(\overline{m}^2_q))
\end{eqnarray}
and substituting it into (\ref{mps}) one can eliminate  the dependence on  
the factorisation scale $\mu_f$ and change it to the $\rm{\overline{MS}}$-scheme running   mass $\overline{m}_q(\overline{m}^2_q)$.
To finalise this section we note that   it is possible to obtain the link between the pole heavy quark mass and    
the $\rm{\overline{\rm{MS}}}$ running  mass through  the subtracted-potential mass. Therefore 
the  explicit expressions for the  
coefficients $a^{(0)}_3$ and  $a^{(1)}_3$  in (\ref{a3}) can  be  related  
with the contributions to the  $\mathcal{O}(a_s^4)$ coefficients  in the 
the ratio of the running and pole heavy quark masses
and may be used 
in future to check the  
results to be discussed 
in the next section.

\section{The four-loop relation between  $\rm{\overline{\rm{MS}}}$ running and pole heavy quark masses.}

Perturbative relation between 
pole and $\rm{\overline{MS}}$ running heavy quark masses contains  the 
discovered in \cite{Beneke:1994sw}, \cite{Bigi:1994em}  
long-distance IR renormalon effects, which result 
from the    
asymptotic structure of the  perturbative QCD  series 
(for discussions of the asymptotic behaviour 
of the perturbative series in the  
renormalised quantum field 
theory models see  \cite{Kazakov:1980rd}). 
However, 
to understand when the asymptotic nature of these expansions is starting 
to manifest itself and when the truncated perturbative series can be used 
it is necessary to study the concrete values of high-order corrections. 
To analyse    these problems we consider the ratios of the defined in the    
$\rm{\overline{MS}}$-scheme running heavy quark masses  and  
the pole heavy quark masses, namely
\begin{eqnarray}
\label{zm}
\frac{\overline{m}_q(M_q^2)}{M_q}=1+\sum\limits_{i=1}^{\infty} z^{(i)}_ma^i_s(M_q^2)~~.
\end{eqnarray}
The coefficients  $z^{(i)}_m$  can be expanded  in powers of $n_l$. 
All coefficients $z^{(i)}_m$ with $1 \leq i \leq 3$ were  evaluated analytically in the  case of $\rm{SU}(N_c)$ colour  gauge group. 
The one-loop correction was found in \cite{Tarrach:1980up}, the two-loop contribution 
was analytically evaluated in  \cite{Gray:1990yh},  the three-loop term is known from 
the analytical and semi-analytical calculations, performed in \cite{Melnikov:2000qh} and \cite{Chetyrkin:1999qi} correspondingly.  
The coefficient  $z^{(4)}_m$ can be presented in the analogous to  equation (\ref{a3}) form:         
\begin{eqnarray}
\label{zm4}
z^{(4)}_m=z^{(40)}_m+z^{(41)}_m n_l+z^{(42)}_m n^2_l+z^{(43)}_m n^3_l 
\end{eqnarray}
The  terms $z^{(43)}_m$ and $z^{(42)}_m$ are known in the 
analytical form from the calculations of  \cite{Lee:2013sx}.
In the case of $\rm{SU}_c(3)$ group of colour symmetry their expressions read:
\begin{eqnarray}
\label{Z43}
z^{(43)}_m&=&\frac{42979}{1119744}+\frac{317\zeta_3}{2592}+\frac{89\pi^2}{3888}+\frac{71\pi^4}{25920}~, \\ \nonumber
z^{(42)}_m&=&-\frac{32420681}{4478976}-\frac{40531\zeta_3}{5184}-\frac{63059\pi^2}{31104}-\frac{103\pi^2\ln 2}{972}+\frac{11\pi^2\ln^2 2}{243}
-\frac{2\pi^2\ln^3 2}{243}-\frac{5\pi^2\zeta_3}{48}+ \\ \label{z42} 
&+&\frac{241\zeta_5}{216}-\frac{30853\pi^4}{466560}-\frac{31\pi^4\ln 2}{9720}+\frac{11\ln^4 2}{486}
-\frac{\ln^5 2}{405}+\frac{44}{81}\rm{Li}_4\left(\frac{1}{2}\right)+\frac{8}{27}\rm{Li}_5\left(\frac{1}{2}\right)
\end{eqnarray}
First two terms in (\ref{zm4})  are not yet known  analytically. However,   
the overall numerical expressions for  $z^{(4)}_m$ at fixed $n_l$ 
were calculated in  \cite{Marquard:2015qpa} with
the help of computer program  FIESTA, created                              
in \cite{Smirnov:2008py}, \cite{Smirnov:2013eza}. The obtained in  
\cite{Marquard:2015qpa} results  read:
\begin{eqnarray}
\label{fixedn}
z^{(4)}_m \bigg\vert_{n_l=3}=-1744.8\pm 21.5 , \;\; z^{(4)}_m \bigg\vert_{n_l=4}=-1267.0\pm 21.5 , \;\; 
 z^{(4)}_m \bigg\vert_{n_l=5}=-859.96\pm 21.5~~~. 
\end{eqnarray}
They  were recently  used by us in  \cite{Kataev:2015gvt}  to get the 
numerical expressions for the analytically unknown terms $z^{(40)}_m$ and  
$z^{(41)}_m$ with the help of 
rigorous optimal least squares (OLS) mathematical method. Bellow we summarise definite steps of this work 
and discuss its  main results. To get the values of {\it two} unknown parameters from the results of (\ref{fixedn}) we 
use the following  overdetermined system of {\it three}  linear equations: 
\begin{eqnarray}
\label{system}
z^{(40)}_m+3z^{(41)}_m=-1371.77 \;, ~~~
z^{(40)}_m+4z^{(41)}_m=-614.68 \;, ~~~
z^{(40)}_m+5z^{(41)}_m=142.32 
\end{eqnarray}
To solve this system by means of the OLS method the following function is introduced  
\begin{eqnarray*}
\Phi(z^{(40)}_m, z^{(41)}_m)=\sum\limits_{k=1}^3 \Delta^2_{l_k}=\sum\limits_{k=1}^3 (z^{(40)}_m+z^{(41)}_m n_{l_k}-y_{l_k})^2
\end{eqnarray*}
where  $\Delta_{l_k}$ are the squared deviations,    
$1\leq k\leq 3$ labels the concrete equations in the system (\ref{system}) and the expressions 
for   $y_{l_k}$ are fixed by the  numbers on 
the  r.h.s. of each  equation from this system.
Its  solutions  $(z^{(40)}_m , z^{(41)}_m)$ are determined by the following 
requirements:
\begin{eqnarray}
\label{FI}
\frac{\partial\Phi}{\partial z^{(40)}_m}=0 , \;\; \frac{\partial \Phi}{\partial z^{(41)}_m}=0 .
\end{eqnarray} 
It is possible to show that their solution  always exists and is  
unique due to the fact that the rank of the matrix, 
composed from  the l.h.s.  of the system, is equal to two and  coincides with the number of unknowns parameters.
This method allows to determine theoretical 
uncertainties of the solutions of Eq.~(\ref{FI}) of the parameters.
Theoretical errors  were  defined in  \cite{Kataev:2015gvt}  as 
\begin{equation}
\label{errors}
\Delta z^{(40)}_m=\frac{\sqrt{\sum\limits_{k=1}^3 n^2_{l_k}}}{\sqrt{3\sum\limits_{k=1}^3 n^2_{l_k}-\left(\sum\limits_{k=1}^3 n_{l_k}\right)^2}}\Delta y_l~~,~~
\Delta z^{(41)}_m=\frac{\sqrt{3}\Delta y_l}{\sqrt{3\sum\limits_{k=1}^3 n^2_{l_k}-\left(\sum\limits_{k=1}^3 n_{l_k}\right)^2}}  
\end{equation}
where  $\Delta y_l=\sigma=21.5$ are  the given in  
\cite{Marquard:2015qpa} uncertainties of the 
results of (\ref{fixedn}).   
Using equations  (\ref{FI}), supplemented by (\ref{errors}), in 
\cite{Marquard:2015qpa}  we found the following expressions 
for two  parameters 
we are  interested in :   
\begin{equation}
\label{results}
z^{(40)}_m(M^2_q)=-3642.9\pm 62.0 , \;\; z^{(41)}_m(M^2_q)=757.05\pm 15.2 .
\end{equation}
Similar expressions  were also obtained   in  \cite{Kiyo:2015ooa} 
with the help of the {\it fitting procedure} of  the numerical results of 
\cite{Marquard:2015qpa} (see Eq.~(\ref{fixedn})), supplemented 
by the calculated  in   \cite{Lee:2013sx}
analytical expressions  for $z^{(43)}_m$ and $z^{(42)}_m$  
and by  
renormalon-based 
large $\beta_0$-representation of the $n_l$-dependence for $z^{(4)}_m$,
fixed  in  \cite{Beneke:1994qe}, \cite{Beneke:1994sw}.
This additional  theoretical input was not 
used by us.  However,  our results from Eq.~(\ref{results}) 
agree very well  with the numbers  
\begin{equation}
\label{fitting}
z^{(40)}_m(M^2_q)=-3643\pm 21.5 , \;\; z^{(41)}_m(M^2_q)=757 \pm 21.5 ,
\end{equation}
which follow from the ones obtained  in \cite{Kiyo:2015ooa}  
for the $\mathcal{O}(a_s^4)$ contributions to the 
inverse ratio  $M_q/\overline{m}_q$,  
where  the running mass  was normalised  at  another scale 
$\mu=\overline{m}_q$. The essential difference is that in     
\cite{Kiyo:2015ooa} 
theoretical errors of the outcomes of their  fits were not   
analysed  separately but  fixed by  the errors of the the numerical 
results,  obtained in  \cite{Marquard:2015qpa}.  

\section{Discussions  of the theoretical errors.}

In our work of \cite{Kataev:2015gvt} we raised the question on the reason 
of the coincidence of the numerical errors in the presented 
in  \cite{Marquard:2015qpa} results of the numerical calculations 
of the $\mathcal{O}(a_s^4)$ corrections  to the relations between running and pole heavy 
masses of $c$, $b$ and $t$-quarks (see Eq.~(\ref{fixedn}). The proposed 
in  \cite{Kataev:2015gvt} explanation of the flavour-independence of the 
numerical error is that this error corresponds to the uncertainty 
of the $n_l$-independent contribution
to (\ref{zm4}), namely of the coefficient  $z^{(40)}_m$. If this explanation is
correct, then the error of the numerical calculations  
of the  coefficient  $z^{(41)}_m$ should be negligibly small.
Therefore,  the   given in 
\cite{Kiyo:2015ooa} errors   are not absolutely correct. 
 Indeed,  they do not satisfy the proposed in  \cite{Kataev:2015gvt} 
explanation of the ``paradox of the coincidence of the numerical errors''. 
The determined in  \cite{Kataev:2015gvt} 
OLS method   error of 
$z^{(40)}_m$-term really dominate over the uncertainty of 
$z^{(41)}_m$ (see (\ref{results}))  and it is   3 times larger 
than the numerical error of this coefficient, which follows 
from the results of \cite{Marquard:2015qpa}.  In spite of this 
positive fact we understand  that the errors obtained   within  
the OLS method    
may be  overestimated (in order to get more definite estimates it is 
necessary to have more than three equations). However, even these results 
indicate that it may be useful to clarify  in more details  
how  the numerical errors of the results of  \cite{Marquard:2015qpa}
were obtained. 
This problem is really important.  Indeed,  the difference in the 
numerical errors discussed above leads to different 
theoretical estimates of the  uncertainties of the   
top-quark pole mass value  (in the case of applications 
of the results of  \cite{Marquard:2015qpa} this error is 0.005~$\rm{GeV}$, 
while if the  results of  \cite{Kataev:2015gvt} are used 
the error is  4  times larger, namely 0.023~$\rm{GeV}$). 
The   analytical evaluation of 
the coefficients  $z^{(40)}_m$ and $z^{(41)}_m$ will    
remove this problem. It is highly desirable to 
get their  analytical expressions.   
These  calculations are realistic and 
already started in  \cite{Lee:2015eva} 
from the creation of 
the special computer program.

\section{Acknowledgements}

We are grateful to V.A. Smirnov for discussions 
of the definite subjects discussed in this talk. One of us (ALK) 
would like to thank the Organisers of ACAT-2016 Workshop in Valparaiso, 
Chilie (18-22 January 2016) for the  invitation and partial financial support. 
This work of ALK was supported by  Russian Science Foundation, 
grant No. 14-22-00161.

\end{document}